\documentclass[amsmath, amsfonts, twocolumn, showpacs, prb]{revtex4}
\usepackage{graphicx}
\usepackage{epsfig}
\usepackage{bm}
\usepackage{dcolumn}
\usepackage{amsmath}
\usepackage{amssymb}
\usepackage{euscript}
\usepackage[varg]{txfonts}

\def\tphi{\tau_{\phi}}
\def\tgl{\tau_{\mathrm{GL}}}
\def\oc{\omega_{H}}
\def\qo{(q,\omega)}
\def\d{\mathrm{d}}
\def\smt{\sigma^{\mathrm{MT}}}
\def\swl{\sigma^{\mathrm{WL}}}
\def\sal{\sigma^{\mathrm{AL}}}

\begin{document}

\title{Magnetoconductivity of low-dimensional disordered conductors \\
at the onset of the superconducting transition}

\author{Alex Levchenko}

\affiliation{Kavli Institute for Theoretical Physics, University of
California, Santa Barbara, California, 93106, USA}

\begin{abstract}
Magnetoconductivity of the disordered two- and three-dimensional
superconductors is addressed at the onset of superconducting
transition. In this regime transport is dominated by the fluctuation
effects and we account for the interaction corrections coming from
the Cooper channel. In contrast to many previous studies we consider
strong magnetic fields and various temperature regimes, which allow
to resolve the existing discrepancies with the experiments.
Specifically, we find saturation of the fluctuations induced
magneto-conductivity for both two- and three-dimensional
superconductors at already moderate magnetic fields and discuss
possible dimensional crossover at the immediate vicinity of the
critical temperature. The surprising observation is that closer to
the transition temperature weaker magnetic field provides the
saturation. It is remarkable also that interaction correction to
magnetoconductivity coming from the Cooper channel, and specifically
the so called Maki-Thompson contribution, remains to be important
even away from the critical region.
\end{abstract}

\date{June 24, 2009}

\pacs{74.25.Fy, 74.40.+k}

\maketitle

Magnetotransport measurements provide a direct access to
localization effects in disordered metals and superconductors. This
is a sensitive technique to probe the nature of electron coherence
and specifically, magnetoresistance experiments opens a way to
determine the electron dephasing time ($\tphi$), which plays a key
role in quantum-interference phenomena. Negative anomalous
magnetoresistance in metals~\cite{HLN} is successfully explained by
the weak-localization effect, see Ref.~\onlinecite{AA} for the
review. Situation becomes more interesting in superconductors since
in the vicinity of the critical temperature transport is dominated
by the superconductive fluctuations~\cite{Larkin-Varlamov} and one
must necessarily account for the interaction corrections coming from
the Cooper channel.~\cite{Aslamazov-Larkin,Maki,Thompson} The
existing theory of the magnetotransport in $d$-dimensional
superconductors~\cite{AA,Larkin,AALK} predicts that \textit{excess}
part of magnetoconductivity
$\delta\sigma_{d}(H)=\sigma_{d}(H)-\sigma_{d}(0)$ for weak
spin-orbit scattering is given by
\begin{eqnarray}\label{sigma-wl+mt}
\delta\sigma_{d}(H)=\delta\swl_{d}(H)+\delta\smt_{d}(H)=\nonumber\\
\frac{e^2}{2\pi^2\hbar}\left(\frac{eH}{\hbar
c}\right)^{\frac{d}{2}-1}\big[1-\beta(T)\big]Y_{d}(\oc\tphi)\,,
\end{eqnarray}
where $\oc=4eDH/c$ is the cyclotron frequency in a disordered
conductor and $D$ is the corresponding diffusion coefficient. The
dimensionality dependent universal function $Y_{d}(x)$ is known from
the localization theory. In the two-dimensional case it is given
by~\cite{AKLL}
\begin{equation}\label{Y2}
Y_{2}(x)=\ln x+\psi\left(\frac{1}{2}+\frac{1}{x}\right)\,,
\end{equation}
with the limiting cases $Y_{2}(x)\approx x^2/24$ for $x\ll1$ and
$Y_{2}(x)\approx\ln x$ for $x\gg1$, where $\psi(x)$ is the digamma
function. In the three-dimensional case~\cite{Kawabata}
\begin{equation}\label{Y3}
Y_{3}(x)=\sum^{\infty}_{n=0}\left[\frac{2}{\sqrt{n+1+\frac{1}{x}}+\sqrt{n+\frac{1}{x}}}-
\frac{1}{\sqrt{n+\frac{1}{2}+\frac{1}{x}}}\right]\,,
\end{equation}
with the limits $Y_{3}(x)\approx x^{3/2}/48$ for $x\ll1$ and
$Y_{3}(x)\approx0.605$ for $x\gg1$.

The first term in the square brackets of Eq.~\eqref{sigma-wl+mt}
corresponds to the conventional weak-localization (WL)
correction.\cite{AKLL,Kawabata} The second term, containing
temperature dependent $\beta(T)$ factor, which is universal
irrespective dimensionality, originates from the interaction
corrections in the Cooper channel, and specifically from the
Maki-Thomspon (MT) diagram.~\cite{Maki,Thompson} It was Larkin's
insightful observation~\cite{Larkin} that interactions with
superconductive fluctuations lead to the same magnetic field
dependence of the conductivity as in the case of weak-localization.
Since $\beta(T)$ is strongly temperature dependent
\begin{eqnarray}
&&\beta(T)=\frac{\pi^2}{6}\frac{1}{\ln^{2}(T/T_{c})}\,,\quad \ln(T/T_c)\gg1\, , \\
&&\beta(T)=\frac{\pi^2}{4}\frac{1}{\ln(T/T_{c})}\,,\quad\,\,\,
\ln(T/T_{c})\ll1\, ,
\end{eqnarray}
where $T_c$ is the critical temperature of a superconductor,
$\delta\smt_{d}(H)$ dominates against $\delta\swl_{d}(H)$ in the
immediate vicinity of the transition when $T-T_{c}\lesssim T_c$. It
is worth emphasizing that MT contribution remains essential even
away from the critical region as well as stays important in the
nonsuperconductive materials, having repulsive interaction in the
Cooper channel, which is in contrast to the Aslamazov-Larkin (AL)
contribution.~\cite{Aslamazov-Larkin} Furthermore, since
$\beta(T)>0$ for any sing of the interaction in the Cooper channel,
Maki-Thompson correction reduces the magneto-conductivity in the
absolute value.

It turns out, however, that in general Eq.~\eqref{sigma-wl+mt} fails
to reproduce experimental observations in both
two-~\cite{Exp-2D-1,Exp-2D-2,Exp-2D-3,Exp-2D-4} and
three-dimensional~\cite{Exp-3D-1,Exp-3D-2,Exp-3D-3,Exp-3D-4,Exp-3D-5}
cases, except for the limit of relatively weak magnetic fields.
Careful experimental analysis revealed that the discrepancy stems
from the Maki-Thomposn part of Eq.~\eqref{sigma-wl+mt}, which ceases
to follow $\delta\smt_{d}(H)=-\beta(T)\delta\swl_{d}(H)$ above the
certain magnetic field. Strictly speaking, validity of
Eq.~\eqref{sigma-wl+mt} relies essentially on the assumptions
\begin{equation}\label{H-assuption}
\oc\lesssim T\ln\frac{T}{T_{c}}\lesssim T\,,\quad\tphi^{-1}\lesssim
T\ln\frac{T}{T_c}\,,
\end{equation}
which set a lower bound for its applicability in the magnetic field,
so that this discrepancy is not surprising and should be
anticipated. In application to the two-dimensional case this problem
and subsequent generalizations were realized in
Refs.~\onlinecite{Santos-Abrahams,Brenig-1,Brenig-2,Reizer}, for the
three-dimensional case there is no theoretical formulation, with the
noticeable exceptions,~\cite{Hikami-Larkin,Dorin} where layered
superconductors were considered.

The purpose of the present work is to give a unified and complete
theory of the magnetotransport in the fluctuating regime of
superconductors. We relax on the assumptions of
Eq.~\eqref{H-assuption} and treat the regime of strong magnetic
field for both two- and three-dimensional cases. The essential
results can be summarized as follows: (i) the excess part of
fluctuation-induced magnetoconductivity, including both
Maki-Thompson and Aslamazov-Larkin contributions, saturates to its
negative, zero-field values at already moderate magnetic fields
$T\ln\frac{T}{T_c}\lesssim\oc\lesssim T$ for any dimensionality.
This fact has clear physical explanation and is supported by all
available experiments. Indeed, magnetic field can be thought as an
effective \textit{depairing factor}, which shifts critical
temperature driving the system away from the transition, thus
suppressing fluctuation effects. At the technical level this happens
because magnetic field enters as the mass of the fluctuation
propagator. (ii) Maki-Thompson magnetoconductivity contribution
dominates against Aslamazov-Larkin for the most of the
experimentally accessible temperatures, except for the immediate
vicinity of the critical temperature. (iii) A surprising and rather
counterintuitive observation is that closer one is to the transition
temperature \textit{weaker} magnetic field leads to
magnetoconductivity saturation, since it is controlled by the ratio
$\oc/T$ and not by $\oc$ itself. This fact is also supported by the
experiments.~\cite{Exp-3D-3,Exp-3D-4} Naively one would expect a
completely different scenario, since proximity to a transition
enhances the lifetime for fluctuating Cooper pairs and thus,
stronger field is required to destroy them. (iv) As temperature is
lowered one may observe a \textit{dimensional crossover} from
three-dimensional case to two dimensional when Ginzburg-Landau
length ($\ell_{\mathrm{GL}}$) exceeds the thickness ($b$) of the
film, namely when
$\ell_{\mathrm{GL}}\sim\sqrt{\frac{D}{T-T_c}}\gtrsim b$. The
indication for this possibility is already seen in the experimental
results of Ref.~\onlinecite{Exp-3D-4}. Another possibility is the
crossover between MT and AL contributions which in principle can be
realized for thicker films or in the layered superconductors due to
their highly anisotropic properties.~\cite{Hikami-Larkin,Dorin}

Quantitatively we find for the Maki-Thompson magnetoconductivity in
the three-dimensional case (hereafter $\hbar=c=k_{B}=1$)
\begin{equation}\label{smt-3}
\delta\smt_{3}(H)=\frac{e^2}{2\pi^2\ell_{H}}\mathcal{B}(T)
\left[\mathcal{Y}_{3}(\oc\tgl)-\mathcal{Y}_{3}(\oc\tphi)\right]\,,
\end{equation}
where magnetic length $\ell_{H}=\sqrt{D/\oc}$ and Ginzburg-Landau
time $\tgl^{-1}=\frac{8T}{\pi}\ln\frac{T}{T_c}$ were introduced. The
universal scaling function reads as
\begin{equation}
\mathcal{Y}_{3}(x)=\int^{+\infty}_{0}\frac{\d
t}{\sqrt{t}}\left[\psi\left(\frac{1}{2}+t+\frac{1}{x}\right)-
\ln\left(t+\frac{1}{x}\right)\right]\,.
\end{equation}
The temperature-dependent factor is defined as
$\mathcal{B}(T)=T\tgl/(1-\tgl/\tphi)$. With the help of well-known
properties of the digamma function and for the experimentally
relevant range of magnetic fields, one finds from Eq.~\eqref{smt-3}
following limiting cases for the magnetoconductivity:
\begin{eqnarray}
&&\hskip-.95cm \delta\smt_{3}(H)\!=\!-\frac{e^2}{96\pi\ell_{\phi}}
\frac{1-\eta^{\frac{3}{2}}}{1-\eta}(T\tgl)(\oc\tphi)^2\,,\,\,
\oc\lesssim\tphi^{-1},\label{smt-3-limit-1}\\
&&\hskip-.95cm
\delta\smt_{3}(H)\!=\!-\frac{e^2}{\pi^2\ell_{\phi}}\frac{T\tgl}{1-\eta}
(\oc\tphi)^{\frac{1}{2}}\,,\,\,\tphi^{-1}\lesssim\oc\lesssim\tgl^{-1},
\label{smt-3-limit-2}\\
&&\hskip-.95cm \delta\smt_{3}(H)\!=\!-\smt_{3}(0)\!+\!\!
\frac{(2\!\!\sqrt{2}\!-\!1)\zeta\big(\frac{3}{2}\big)e^{2}}{4\pi\ell_{T}}
\!\sqrt{\frac{T}{\oc}}\,,\,\,
\oc\gtrsim\tgl^{-1}.\label{smt-3-limit-3}
\end{eqnarray}
Here $\zeta(x)$ is the Riemann zeta function,
$\ell_{\phi}=\sqrt{D\tphi}$ and $\ell_{T}=\sqrt{D/T}$ are dephasing
and thermal lengths respectively, and we introduced
$\eta=\tgl/\tphi$ for compactness. These asymptotes are valid as
long as $\tphi>\tgl$. In the opposite limit one has to interchange
$\tphi\rightleftarrows\tgl$. One sees from
Eqs.~\eqref{smt-3-limit-1}--\eqref{smt-3-limit-3} that excess part
of the magnetoconductivity goes through the series of crossovers
$\delta\smt_{3}(H)\propto H^{2}\to\sqrt{H}\to\mathrm{const}$, until
it saturates to its negative and magnetic-field independent value
\begin{equation}\label{smt-3-H0}
\smt_{3}(0)=\frac{e^2}{\pi\ell_{T}}\frac{\sqrt{T\tgl}}{1+\sqrt{\tgl/\tphi}}\,.
\end{equation}
\begin{figure}[t!]
\begin{center}\includegraphics[width=8cm]{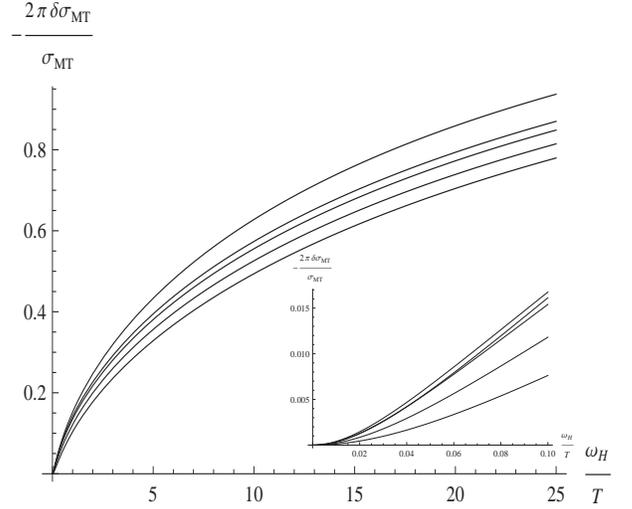}\end{center}
\caption{Normalized Maki-Thompson magnetoconductivity for 3$d$ case
calculated at $T=1.8,3.5,4.4,6.0,8.0\,$K (top to bottom) with the
corresponding dephasing times
$\tphi=(2.24,1.42,1.05,0.58,0.28)\times10^{-10}$s. The material
parameters are $D=6.37\times10^{-4}$m$^{2}$/s and $T_c=123$mK that
correspond to Ref.~\onlinecite{Exp-3D-3}. \label{Fig}}
\end{figure}
It is worth emphasizing that Eq.~\eqref{smt-3-limit-1}, taken at
$\eta\to0$, can be recovered from Eq.~\eqref{sigma-wl+mt} in the
limit when $\oc\lesssim\tphi^{-1}$, with the help of the approximate
form of $Y_{3}(x)$ function [Eq.~\eqref{Y3}], as it should be of
course. The saturation region is \textit{not} captured by
Eq.~\eqref{sigma-wl+mt}, but recovered correctly
[Eq.~\eqref{smt-3-limit-3}] within generalized formulation of MT
magnetoconductivity. To facilitate the comparison between the theory
[Eq.~\eqref{smt-3}] and
experiments~\cite{Exp-3D-1,Exp-3D-2,Exp-3D-3,Exp-3D-4} we plot on
the Fig.~\ref{Fig} the MT magnetoconductivity at different
temperatures for the material parameters taken from
Ref.~\onlinecite{Exp-3D-3}. The inset plot in Fig.~\ref{Fig}
emphasizes quadratic magnetic-field dependence of
$\delta\smt_{3}(H)$ at the lowest fields [see
Eq.~\eqref{smt-3-limit-1}].

For the two-dimensional case magnetoconductivity is determined by
the following expression~\cite{Santos-Abrahams,Brenig-2,Reizer}
\begin{equation}\label{smt-2}
\delta\smt_{2}(H)=\frac{e^2}{\pi}\mathcal{B}(T)
\left[Y_{2}(\oc\tgl)-Y_{2}(\oc\tphi)\right]\,,
\end{equation}
where $Y_{2}(x)$ is defined by Eq.~\eqref{Y2}. For the same range of
magnetic fields as in
Eqs.~\eqref{smt-3-limit-1}--\eqref{smt-3-limit-3} one finds from
Eq.~\eqref{smt-2}
\begin{eqnarray}
&&\hskip-.9cm
\delta\smt_{2}(H)=-\frac{e^2}{24\pi}\left[1+\frac{\tgl}{\tphi}\right]
(T\tgl)(\oc\tphi)^2\,,\,\,\oc\lesssim\tphi^{-1}\!,\label{smt-2-limit-1}\\
&&\hskip-.9cm
\delta\smt_{2}(H)=-\frac{e^2}{\pi}\frac{T\tgl}{1-\frac{\tgl}{\tphi}}
\ln\frac{\oc\tphi}{4e^{\gamma_{E}}}\,,\,\,
\tphi^{-1}\lesssim\oc\lesssim\tgl^{-1},\label{smt-2-limit-2}\\
&&\hskip-.9cm \delta\smt_{2}(H)=-\smt_{2}(0)+\frac{\pi
e^2}{2}\frac{T}{\oc}\,,\,\,\oc\gtrsim\tgl^{-1},\label{smt-2-limit-3}
\end{eqnarray}
where $\gamma_{E}=0.57..$ is the Euler constant. Similarly to the
three-dimensional case MT magnetoconductivity saturates trough the
series of crossovers $\delta\smt_{2}(H)\propto H^2\to\ln
H\to\mathrm{const}$, to its field independent value determined
by~\cite{Thompson}
\begin{equation}
\smt_{2}(0)=\frac{e^2}{\pi}\frac{T\tgl}{1-\tgl/\tphi}\ln\frac{\tphi}{\tgl}\,.
\end{equation}
By comparing Eq.~\eqref{smt-3-limit-1} to Eq.~\eqref{smt-2-limit-1}
one concludes that quadratic field dependence,
$\delta\smt_{d}(H)\propto H^2$, at the lowest fields,
$\oc\lesssim\tphi^{-1}$, is apparently universal, in agreement with
Eq.~\eqref{sigma-wl+mt}, while magnetoconductivity saturation in the
two-dimensional case is stronger then in the three dimensions.

At this point we discuss the role of Aslamazov-Larkin contribution
to the magnetoconductivity and compare it to $\delta\smt_{d}(H)$. In
the two-dimensional case we find
\begin{eqnarray}\label{sal-2}
&&\sal_{2}(H)=\frac{2e^2}{\pi}(T\tgl)\mathcal{H}_{2}(\oc\tgl)\,,\\
&&\mathcal{H}_{2}(x)=\frac{1}{x}\left[1-\frac{2}{x}
\left[\psi\left(1+\frac{1}{x}\right)-
\psi\left(\frac{1}{2}+\frac{1}{x}\right)\right]\right]\,.
\end{eqnarray}
With the help of the asymptotic form of $\mathcal{H}_{2}$ function
at zero field, $H\to0$, where $\mathcal{H}_{2}(x\to0)\to1/4$, one
recovers from Eq.~\eqref{sal-2} famous
result~\cite{Aslamazov-Larkin}
\begin{equation}\label{sal-2-H0}
\sal_{2}(0)=\frac{e^2}{16}\frac{1}{\ln(T/T_c)}\,.
\end{equation}
Subtracting now $\sal_{2}(0)$ from Eq.~\eqref{sal-2} for the two
limiting cases of low,
$\mathcal{H}_{2}(x)\approx\frac{1}{4}(1-\frac{x^2}{8})$, when
$x\ll1$, and high, $\mathcal{H}_{2}(x)\approx1/x$, when $x\gg1$,
magnetic fields one finds:
\begin{eqnarray}
&&\delta\sal_{2}(H)=-\frac{e^2}{16\pi}(T\tgl)(\oc\tgl)^2\,,\,\,
\oc\lesssim\tgl^{-1}\,,\label{sal-2-limit-1}\\
&&\delta\sal_{2}(H)=-\sal_{2}(0)+\frac{2e^2}{\pi}\frac{T}{\oc}\,,\,\,
\oc\gtrsim\tgl^{-1}\,,\label{sal-2-limit-2}
\end{eqnarray}
which agrees also with the earlier
results.~\cite{Hikami-Larkin,Abrahams,Redi} Similarly to
Eqs.~\eqref{smt-3-limit-1} and \eqref{smt-2-limit-1} the low-field
Aslamazov-Larkin magnetoconductivity is universal and scales
quadratically with $\oc$. It also saturates to the field independent
value [Eq.~\eqref{sal-2-H0}] at $\oc\gtrsim\tgl^{-1}$, having the
same $\sim1/H$ correction as in Eq.~\eqref{smt-2-limit-3}. However,
if one compares the magnitude of the MT and AL contributions, for
example at $\oc\sim\tgl^{-1}$, then it is easy to see from
Eqs.~\eqref{smt-2-limit-2} and \eqref{sal-2-limit-1} that
$\delta\smt_{2}(H)$ dominates against $\delta\sal_{2}(H)$ by the
logarithmic factor $\ln(\tphi/\tgl)$ and this tendency persists for
the smaller fields. Although $\ln(\tphi/\tgl)$ depends on
temperature, it actually stays practically constant,
$\ln(\tphi/\tgl)\sim5$, at the experimentally addressed range of
temperatures, 1K$\lesssim T\lesssim$10K in most of the measurements,
see for example Refs.~\onlinecite{Exp-3D-3} and
\onlinecite{Exp-3D-4}. In the three-dimensional case expression
similar to Eq.~\eqref{sal-2} can be derived,~\cite{Hikami-Larkin}
which brings however the same conclusion about the relative
importance of $\delta\sal_{3}(H)$ when compared to
$\delta\smt_{3}(H)$ [Eq.~\eqref{smt-3}]. It should be emphasized
that situation may be different if $\tgl>\tphi$, which may happen in
the layered superconductors. For this case $\delta\sal_{d}(H)$
dominates the magnetotransport in the vicinity of the critical
temperature.~\cite{Hikami-Larkin,Dorin}

In the remaining part of the paper we outline the essential steps
needed to derive Eqs.~\eqref{smt-3}--\eqref{sal-2}. Within the
linear response Keldysh technique, which is proven to be very
effective tool in application to the transport problems of
fluctuating superconductors,~\cite{Reizer,LK} Maki-Thompson
conductivity correction is determined by the following expression
\begin{eqnarray}\label{smt-def}
\smt_{d}=\frac{e^2D}{2\pi}\sum_{q}\iint^{+\infty}_{-\infty}
\frac{\d\epsilon\d\omega}{\cosh^{2}\frac{\epsilon}{2T}}\,
\mathrm{Im}\big[L^{R}\qo\big]\nonumber\\
\big|C^{R}(q,2\epsilon+\omega)\big|^{2}
\left[\coth\frac{\omega}{2T}-\tanh\frac{\epsilon+\omega}{2T}\right]\,.
\end{eqnarray}
Here interaction propagator is given by
$L^{R}\qo=\left[\ln\frac{T}{T_c}+\psi\left(\frac{Dq^2-i\omega}{4\pi
T}+\frac{1}{2}\right)-\psi\left(\frac{1}{2}\right)\right]^{-1}$,
while $C^{R}\qo=\left[Dq^2-i\omega+\tphi^{-1}\right]^{-1}$ stands
for the Cooperon. In the three-dimensional case with magnetic field
pointed along the $z$ axes one has $Dq^2\to Dq^{2}_{z}+\oc(n+1/2)$
and momentum summation in Eq.~\eqref{smt-def} is performed as
$\sum_{q}\to\frac{\oc}{4\pi D}\int^{+\infty}_{-\infty}\frac{\d
q_{z}}{2\pi}\sum^{\infty}_{n=0}$, where the prefactor conventionally
accounts for the degeneracy in the position of Landau orbit. Passing
to the dimensionless units $x=Dq^{2}_{z}/T$, $y=\omega/T$,
$z=\epsilon/T$, and $w_{n}=\frac{\oc}{T}(n+1/2)$ Eq.~\eqref{smt-def}
can be reduced to
\begin{eqnarray}\label{smt-3-integral}
&&\smt_{3}(H)=
\frac{e^2}{2\pi^4\ell_{T}}\frac{\oc}{T}\sum^{\infty}_{n=0}
\int^{+\infty}_{0}\frac{\d
x}{\sqrt{x}}\iint^{+\infty}_{-\infty}\!\!\frac{\d z\d
y}{\cosh^2\frac{z}{2}}\nonumber\\
&&\frac{y\big[\coth\frac{y}{2}-\tanh\frac{z+y}{2}\big]}
{\big[\big(x+w_{n}+\frac{1}{T\tgl}\big)^2+y^2\big]
\big[\big(x+w_{n}+\frac{1}{T\tphi}\big)^2+(2z+y)^2\big]}\,,\qquad
\end{eqnarray}
where we expanded interaction propagator $L^{R}\qo$ at small
frequencies and momenta, assuming
$\mathrm{max}\{Dq^2,\omega\}\ll4\pi T$. One sees from
Eq.~\eqref{smt-3-integral} that the relevant range for $z$
integration is set by $z\sim1$, whereas the width of the Cooperon is
determined by $\mathrm{max}\{x,w_{n},(T\tphi)^{-1}\}\ll1$. At the
end, this condition limits applicability of Eq.~\eqref{smt-3} to
magnetic fields not exceeding $\oc\lesssim4\pi T$, which is still
sufficient to explain the magnetoconductivity saturation happening
at $\oc\sim\tgl^{-1}\ll T$. Under this assumption one is allowed to
approximate
$\big|C^{R}(x+w_{n},2z+y)\big|^{2}\approx\frac{\pi}{x+w_{n}+\frac{1}{T\tphi}}\delta(2z+y)$
in Eq.~\eqref{smt-3-integral}. Integration over $z$ becomes
immediate and gives
\begin{eqnarray}\label{smt-3-integral-1}
\smt_{3}(H)=\frac{e^2}{4\pi^3\ell_{T}}\frac{\oc}{T}\sum^{\infty}_{n=0}
\int^{+\infty}_{0}\frac{\d
x}{\sqrt{x}\big[x+w_{n}+\frac{1}{T\tphi}\big]}\nonumber\\
\int^{+\infty}_{-\infty}\frac{\d
y}{\cosh^{2}\frac{y}{4}}\frac{y\big[\coth\frac{y}{2}-\tanh\frac{y}{4}\big]}
{\big[\big(x+w_{n}+\frac{1}{T\tgl}\big)^2+y^2\big]}\,.
\end{eqnarray}
The $y$ integration can be completed with the same line of reasoning
as in the $z$ case, assuming that
$\mathrm{max}\{x,w_{n},(T\tgl)^{-1}\}\ll1$, which is consistent with
the previous step, and gives a factor $2\pi/(x+w_{n}+1/T\tgl)$.
After that step summation over $n$ is straightforward with the help
of the digamma function. Rescaling also $x\to(\oc/T)t$ one obtains
from Eq.~\eqref{smt-3-integral-1}
\begin{eqnarray}\label{smt-3-integral-2}
\smt_{3}(H)=\frac{e^2}{2\pi^2\ell_{H}}\mathcal{B}(T)\int^{+\infty}_{0}\frac{\d
t}{\sqrt{t}}\left[\psi\left(\frac{1}{2}+t+\frac{1}{\oc\tgl}\right)\right.\nonumber\\
\left.-\psi\left(\frac{1}{2}+t+\frac{1}{\oc\tphi}\right)\right]\,.
\end{eqnarray}
To recover Eq.~\eqref{smt-3-H0} from Eq.~\eqref{smt-3-integral-2} at
zero magnetic field one uses following asymptote
$\psi\big(\frac{1}{2}+t+\frac{1}{x}\big)\stackrel{x\to0}{\longrightarrow}\ln\big(t+\frac{1}{x}\big)$
and an integral identity $\int^{\infty}_{0}\frac{\d
t}{\sqrt{t}}\ln\big(\frac{t+a}{t+b}\big)=2\pi\big[\sqrt{a}-\sqrt{b}\big]$.
As the final step one subtracts Eq.~\eqref{smt-3-H0} from
Eq.~\eqref{smt-3-integral-2} and arrives at our major result given
by Eq.~\eqref{smt-3}. Corresponding calculations for the
two-dimensional case [Eq.~\eqref{smt-2}] as well as derivation of
Aslamazov-Larkin contribution [Eq.~\eqref{sal-2}] are completely
analogous.

In conclusion we have suggested the complete theoretical description
of the magnetotransport in fluctuating regime of superconductors of
different dimensionality. Interaction corrections in the Cooper
channel play the dominant role and are governed by the Maki-Thompson
contribution. Sufficiently strong magnetic field suppresses
fluctuation effects completely and magnetoconductivity is determined
then by the weak-localization effect. At the immediate vicinity of
the critical temperature Aslamazov-Larkin correction may become more
important and one may observe MT$\to$AL or dimensional crossovers.
These theoretical results are in good agreement with the
experimental
observations.~\cite{Exp-3D-1,Exp-3D-2,Exp-3D-3,Exp-3D-4}

This research was supported in part by the National Science
Foundation under Grants No. PHY05-51164, No. DMR-0405212, and No.
DMR-0804266.


\end{document}